\documentstyle[11pt,newpasp,twoside,epsf]{article}
\def\edcomment#1{\iffalse\marginpar{\raggedright\sl#1\/}\else\relax\fi}
\reversemarginpar

\begin{document}
\title{AMiBA, XMM, and Cluster Surveys}
 \author{Haida Liang}
\affil{Physics Dept., University of Bristol, Tyndall Ave., Bristol BS8 1TL, UK}

\begin{abstract}
The Array for Microwave Background Anisotropy (AMiBA) is an
interferometric array of 19 dishes co-mounted on a steerable
platform and operating at 95\,GHz. One of the main scientific aims of
AMiBA is to conduct cluster surveys using the Sunyaev-Zel'dovich (SZ)
effect. Here we explore the potential of AMiBA as a tailor-made SZ
instrument for the study of cluster physics and cosmology via cluster
surveys out to the epoch of cluster formation. In particular, we
explore the potential of combining AMiBA cluster surveys with the
XMM-LSS (Large Scale Structure) survey.

\end{abstract}

\section{Introduction}

AMiBA (Array for Microwave Background Anisotropy) is a telescope
funded in Jan. 2000 by the Ministry of Education and National Science
Council in Taiwan. The project is a collaboration between ASIAA in
Taiwan, the National Taiwan University and several international
groups. One of the main scientific aims of the telescope is to
conduct surveys of clusters of galaxies using the Sunyaev-Zel'dovich
effect.

The Sunyaev-Zel'dovich (SZ) effect is the distortion of the Cosmic
Microwave Background (CMB) along the line of sight towards a hot
cluster of galaxies because of the inverse Compton scattering of the
CMB photons by electrons in the cluster medium (Sunyaev \& Zel'dovich
1972). It is proportional to the integral of pressure along the line
of sight. The SZ effect is redshift ($z$) independent because the
$(1+z)^{4}$ dimming of surface brightness is compensated by the
$(1+z)^{4}$ increase in the CMB energy density.  The
redshift-independent nature of the SZ effect makes it good for finding
very distant clusters where conventional optical and X-ray methods
fail because of strong $(1+z)^{4}$ dimming of surface brightness.  A
flux limited catalogue of clusters detected through the SZ effect is
effectively a mass-limited catalogue of high-redshift clusters.

The importance of the SZ effect as a probe for cluster physics and
cosmology has been review by Rephaeli (1995) and Birkinshaw
(1999). Here we will only give a brief summary:
\begin{itemize}
\item $H_{0}$\\ The angular diameter distance to a cluster can be
estimated from a combined analysis of SZ profile, X-ray surface
brightness and temperature profile. A cluster Hubble diagram of
angular diameter distance versus redshift can be constructed from a
sample of clusters with well measured SZ, X-ray surface brightness and
temperature profiles. The Hubble constant can be deduced from the low
redshift ($z<0.2$) linear part of the cluster Hubble diagram. A Hubble
constant thus obtained offers an independent check on conventional
methods based on standard candles in the local Universe.
\item $\Lambda_{0}$\\ The high redshift part of the cluster Hubble
diagram provides a constraint on $\Lambda_{0}-\Omega_{m}$ which is
roughly orthogonal to that imposed by the CMB primary anisotropy
results. Hence $\Lambda_{0}$ can be determined in combination with the
CMB anisotropy results in much the same fashion as the supernova Ia
results. The CMB primary anisotropy probes the CMB at $z\sim 1000$,
but both the supernova Ia and cluster methods probe the CMB at
$z<5$, thus providing independent constraints.
\item Cluster formation and evolution\\ Since a flux-limited catalogue
of clusters detected by the SZ effect corresponds to a mass-limited
catalogue of high redshift clusters, we can probe the epoch of cluster
formation. A mass-limited cluster abundance as a function of redshift
is a direct measure of cluster evolution.
\item $\sigma_{8}$ and $\Omega_{m}$ \\ The cluster abundance as a
function of redshift is sensitive to the normalisation of the
power spectrum $\sigma_{8}$, and to a lesser extent the matter density
$\Omega_{m}$ (e.g. Fan \& Chiueh 2001, Holder et al. 2000).
\item Baryon fraction\\ The ratio of the gas mass deduced from the
integrated SZ effect and the mass deduced from gravitational lensing
provides a reliable lower limit to the baryonic fraction of a
cluster. If clusters are fair samples of the Universe, this is one of
the most precise and model-independent methods for determining the
baryon fraction of the Universe.
\item Cluster astrophysics\\ Since the SZ effect has a different
dependency on the electron density and temperature than to the
X-rays from a cluster, it provides extra constraints on gas properties, such as
clumpiness of the gas, if we assume that the cosmology is known.

\end{itemize}

Up to the present time, work on the SZ effect has been limited by the
existing technology: there are only about 20~clusters with reported SZ
effects after about 20~years of effort. Most of the early work was
based on single dish radio observations.  However, demonstrations over
the past eight years have shown that interferometers are effective at
measuring SZ effects (Jones et al. 1993; Carlstrom et al. 1996). The
advantages of interferometers over single dishes are: 1) less effect
from the atmosphere since interferometers will resolve out most of the
smooth atmosphere; 2) no correlated ground spill-over effects and less effect
from man-made interference since interferometers register only
correlated signals; 3) the ability to make measurements at different
spatial scales simultaneously and thus allowing efficient mapping and
the subtraction of discrete radio sources. Conventional
interferometers such as the VLA are generally designed to achieve high
resolution and high point source sensitivity. However, the SZ effect
from clusters are diffuse and extended, and it is {\em not} the point
source sensitivity but the brightness sensitivity that is need for
detection.  A telescope is most sensitive at detecting a cluster when
its beam is well matched to the size of the cluster. Conventional
interferometers tend to resolve out the diffuse SZ effect. So far even the
``adapted'' SZ instruments such as the Ryle Telescope at 15\,GHz, and
BIMA and OVRO at 30\,GHz are capable of detecting the $\sim 50$
brightest clusters in the sky. A purpose built array of a large number
of small dishes is necessary for blank SZ cluster surveys. It is on
this basis that the AMiBA project was initiated. AMiBA is the first
fully funded SZ effect specific array. Two other arrays: SZA at
30\,GHz, and AMI at 15\,GHz have also been funded recently.  The
availability of the XMM and Chandra orbiting X-ray observatories and
the new generation of purpose built SZ telescopes, provides an
exciting opportunity for achieving some of the science goals described
above.

\section{AMiBA as a SZ instrument}

AMiBA is designed to have 19 dishes co-mounted on a steerable
platform, operating at 95\,GHz with a bandwidth of 20\,GHz and full
polarisation capability. There will be 2 sets of dishes, 1.2\,m and
0.3\,m for the SZ cluster survey and CMB polarisation anisotropy
measurements. The expected system temperature is 75\,K,
so that AMiBA will achieve a flux density sensitivity of 1.4\,mJy/beam
in 1 hour with the 1.2\,m dishes assuming a telescope efficiency of 0.6.

The operating frequency of 95\,GHz was chosen because it is the
frequency window believed to have the least confusion from
astronomical sources, and it is the lowest frequency at which enough
antennas of the right dish size toned to the optimum angular scale for
cluster surveys can be fit on a platform (e.g. a 30\,GHz array would
require dish size $>2$\,m for efficient cluster surveys, hence the
required platform size would be too large for it to be practical).
The advantages of platform arrays are the reduced cost and complexity
in the correlator design and the ability to close pack the dishes to
achieve the best brightness sensitivity without shadowing (see
Fig.~1). At 95\,GHz, the contribution from discrete astronomical
sources as extrapolated from low frequency (8\,GHz and 30\,GHz) source
counts is negligible. We have investigated the statistical confusion
noise from faint SZ sources below the detection limit through mock
observations of simulated SZ sky (e.g. Da Silva et al. 2000), and
found that the contribution is negligible even for a 40hr deep
integration.  The main source of confusion is expected to be from the
primary CMB anisotropy itself (see Subrahmanyan this volume), however
since the power spectrum of the CMB drops exponentially with
decreasing scale, it can be effectively filtered out in the $l$-space
without significant loss of sensitivity.

\begin{figure}
\label{shad}
\epsfysize 200pt \epsfbox{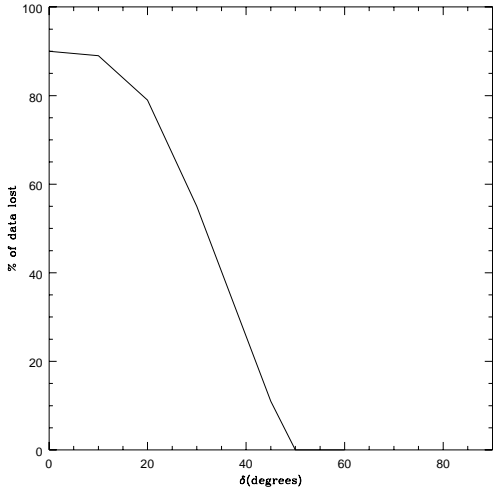}
\caption{ The percentage loss of data due to shadowing as a function
of declination for a BIMA D-array. It is based on an observation over
hour-angle of -3h to +3h. Unlike BIMA, AMiBA is a platform array which
avoids shadowing.}
\end{figure}

Telescopes such as MAP and the CBI have relatively large beam ($\sim
10^{'}$) and are only suitable for shallow surveys of nearby, massive
clusters. The class of ``adapted'' SZ telescopes such as Ryle, BIMA
and OVRO have small beam size of $\sim 1^{'}$ which is good for
mapping the $\sim 50$ brightest clusters, but not suitable for blank
surveys. The new generation of dedicated SZ survey telescopes, AMiBA,
SZA and AMI are designed such that they are most suitable for deep
surveys of low mass clusters. The 3 telescopes complement each other
in frequency and have comparable sensitivity and beam size.

\section{SZ cluster survey strategy}

We plan to survey clusters in 3 modes (an observing efficiency of
75\% is assumed in the following calculations):
\begin{itemize}
\item Deep\\ This survey is geared towards detecting low-mass and
high-redshift clusters to probe the epoch of cluster formation and to
measure the curvature of the Universe. We plan to survey 5\,deg$^{2}$ over $\sim 7$ months, i.e. on average 20\,hrs per field. This
survey is expected to be complete to a mass limit of $\sim 1.5\times
10^{14}$\,M$_{\odot}$. Figure~2 shows
a simulated AMiBA deep survey map.
\item Medium\\ This survey is matched in survey area to the XMM-LSS
survey. We plan to cover a total of 70 sq. degrees in $\sim 7$ months
(average $\sim 1.5$\,hrs per field), reaching a mass completeness
limit of $\sim 3.5\times 10^{14}$\,M$_{\odot}$.
\item Shallow\\ A total of 175 sq. degrees will be surveyed in $\sim
6$ months (average of $\sim 30$\,mins per field) down to a mass limit of
$5\times 10^{14}$\,M$_{\odot}$.
\end{itemize}

We calculate the mass limits using two methods.  The most widely used
method assumes that clusters follow self-similar scaling. Here we use
the self-similar scaling $M_{200}-T$ relation (Eke et al. 1996)
normalised to a relaxed nearby cluster (Mohr et al. 1999), assume a
constant gas fraction of $0.2 h^{-3/2}_{50}$ and a $\Lambda$CDM
universe of $H_{0}=65$\,km\,s$^{-1}$\,Mpc$^{-1}$, $\Omega_{m}=0.3$,
$\Lambda_{0}=0.7$. However, self-similar scaling with constant gas
fraction ($f_{gas}$) is not consistent with observed clusters, for
example based on the evidence of the $L_{x}-T$ relation (Arnaud \&
Evrard 1999), and the $M-T$ relation for nearby clusters (Mohr et
al. 1999). Hence, for comparison we have also calculated the mass
limits by assuming that the locally determined $M-T$ and $f_{gas}-T$
relations (Mohr et al. 1999) are applicable to clusters at all
redshifts. In this case, we have assumed for convenience an
Einstein-de Sitter universe with $H_{0}=50$\,km\,s$^{-1}$\,Mpc$^{-1}$
($\Omega_{m}=1.0$, $\Lambda_{0}=0$).  The mass limits for the surveys
are shown as a function of redshift in Fig.~3 for both sets of
assumptions.

\begin{figure}
\plottwo{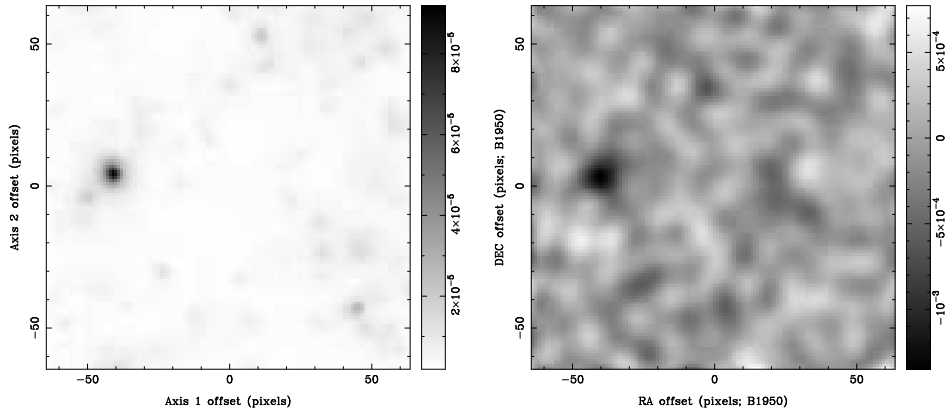}{lcdm04.amiba.vps}
\caption{LEFT: A simulated piece of SZ sky from Da Silva et al. (2000). RIGHT:
Mock AMiBA observations of the same piece of sky in a deep survey
showing the detection of a cluster at $z\sim 2$.}
\label{sim}
\end{figure}

\begin{figure}
\plottwo{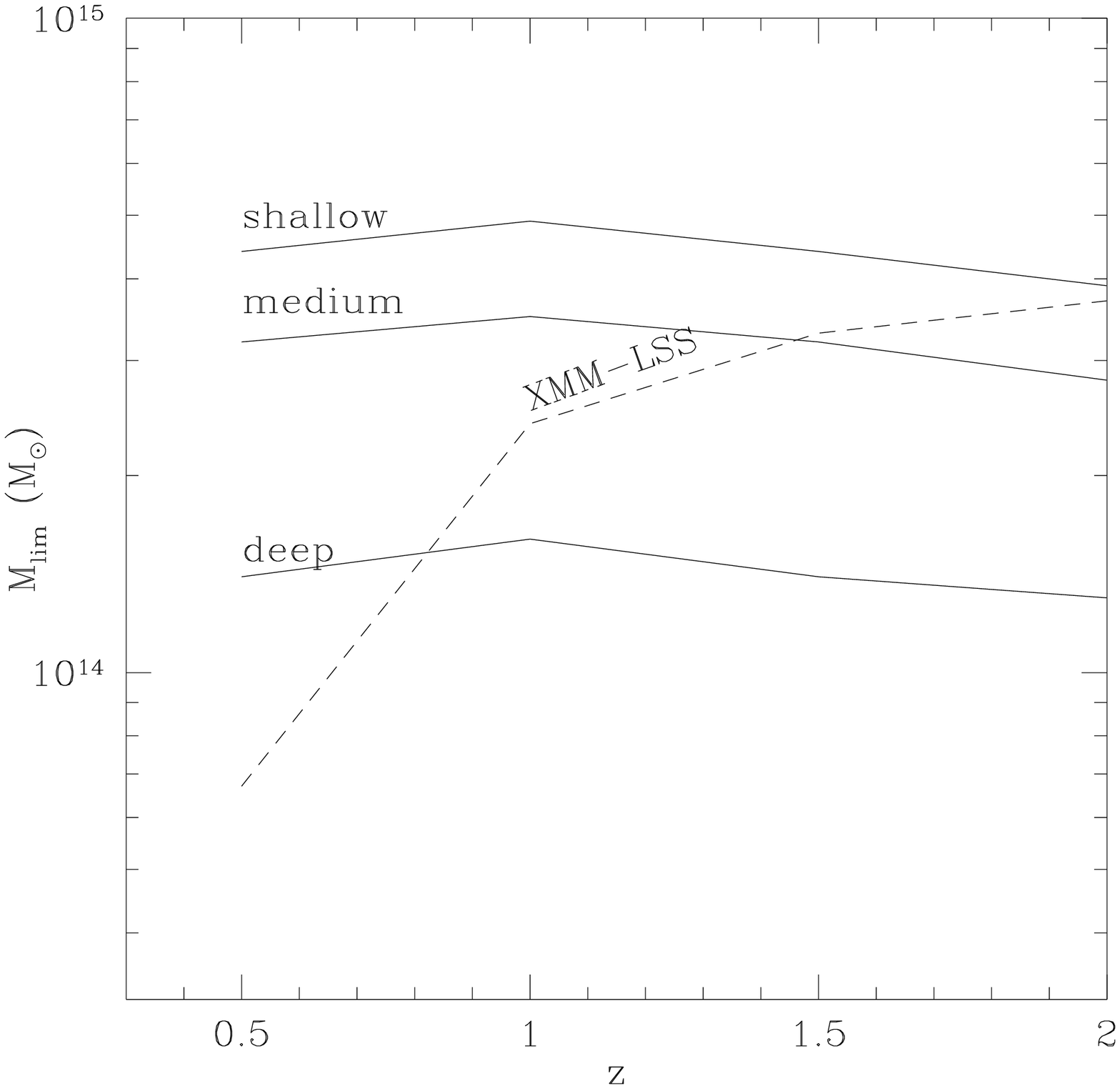}{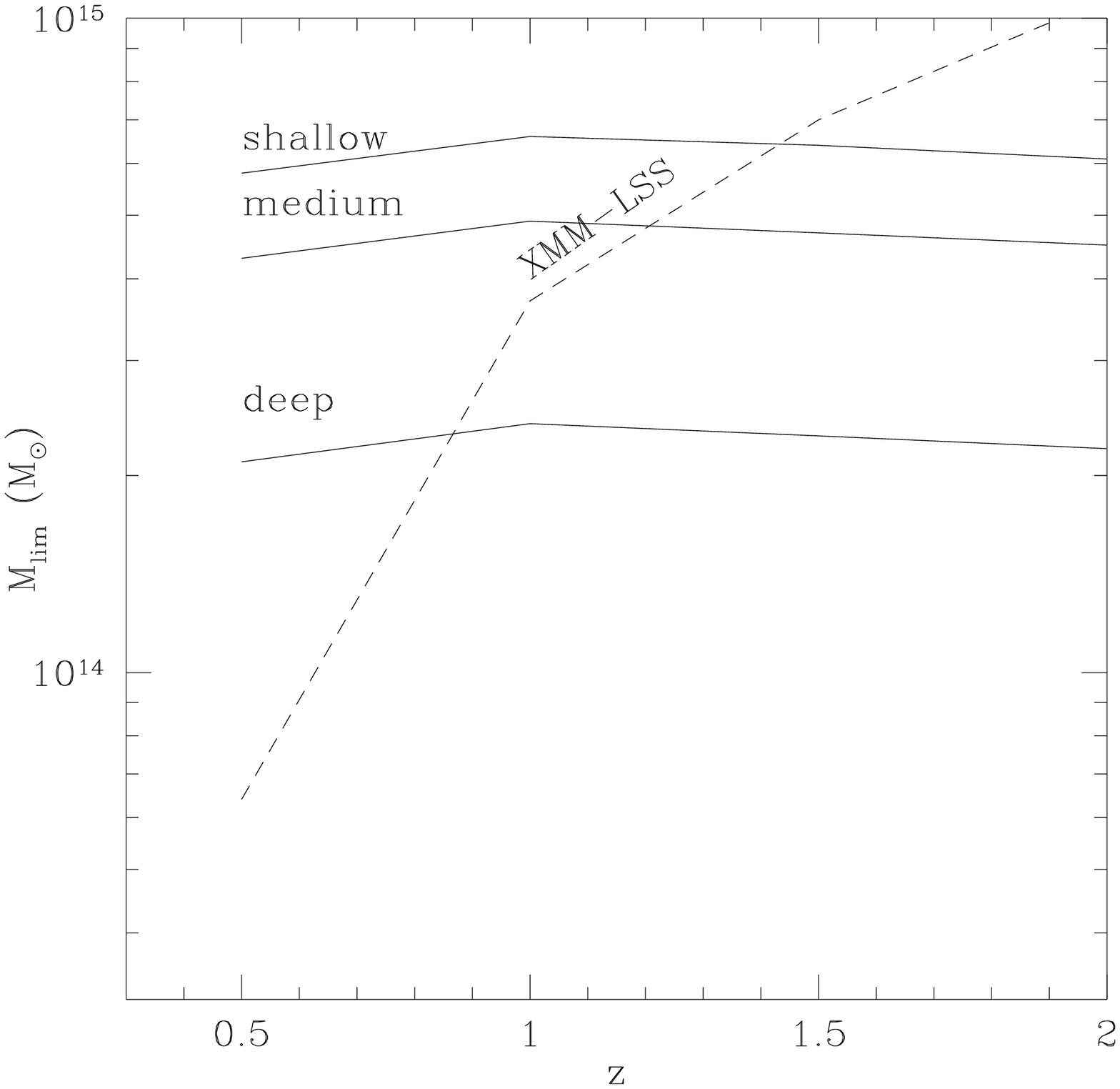}
\caption{The minimum mass ($M_{200}$) for a cluster to be detected by
AMiBA as a function of redshift for deep, medium and shallow surveys
described in the text (solid curves), compared to the $M_{200}$ mass
limit for the XMM-LSS survey in dashed curve. For both sets of
surveys, $M_{lim}$ corresponds to a $5\sigma$ detection
limit. LEFT: Self similar scaling is assumed for a $\Lambda$CDM universe of
$H_{0}=65$\,km\,s$^{-1}$\,Mpc$^{-1}$, $\Omega_{m}=0.3$,
$\Lambda_{0}=0.7$. RIGHT: The $M-T$ relation derived for nearby
clusters (Mohr et al. 1999) is assumed to apply to clusters at all
redshifts, and an Einstein-de Sitter universe of
$H_{0}=50$\,km\,s$^{-1}$\,Mpc$^{-1}$, $\Omega_{m}=1.0$,
$\Lambda_{0}=0$ is assumed. }
\label{mass}
\end{figure}

\section{A complementary X-ray survey -- XMM LSS survey}
The XMM LSS survey (Large Scale Structure survey) aims to map the
large scale structure of the Universe out to redshift $z\sim 1-2$ using
cluster and quasar populations (Pierre et al. 2000,
2001, XMM-LSS WEB page\footnote{http://vela.astro.ulg.ac.be/themes/spatial/xmm/LSS/index\_e.html}). The
survey involves deep XMM mapping down to $5\times
10^{-15}$\,ergs\,cm$^{-2}$\,s$^{-1}$ (0.5-2.0\,keV) of a
$8^{\circ}\times 8^{\circ}$ field near declination zero with follow-up
optical identifications. It is expected to find 300 sources per square
degree, out of which 15-20 are expected to be clusters, 200 QSOs, and
the remainder, stars and galaxies.  The optical follow-up includes
wide-field imaging with MegaCam at CFHT (the XMM-LSS is the priority
target of the CFH Legacy Survey\footnote{
http://cdsweb.u-strasbg.fr:2001/Instruments/Imaging/Megacam/MSWG/forum.html})
and a spectroscopic survey with VIRMOS at VLT. The optical
spectroscopy will enable the construction of the cluster redshift
distribution and thus, for the first time, the computation of the
cluster correlation function in the $0.5<z<1$ interval. The wide-field
imaging with MegaCam will also provide maps of mass distribution from
weak lensing analysis.  Expected cosmological constraints from the
XMM-LSS cluster survey are discussed in detail by Refregier,
Valtchanov \& Pierre (2001).

By surveying the same region with AMiBA, we can target well-matched
clusters to construct a cluster Hubble diagram, and probe cluster
physics and the baryon fraction by combining SZ measurements with
X-rays and lensing analysis. The XMM-LSS survey expects to find a
total of $\sim 50$ clusters (mostly $z<0.5$) for which it will obtain
temperature measurements without further X-ray follow-ups.

While optical and X-ray surveys are efficient at finding a large
number of clusters, i.e. they have a high rate of detection, they do
not produce a statistically unbiased sample of clusters unlike the SZ
effect where a flux limited sample is mass-limited at high
redshift. Figure~3 shows the mass limit as a function of redshift for
the AMiBA surveys as well as the XMM-LSS survey. The AMiBA deep survey
is more sensitive than the XMM-LSS survey at finding $z>0.8$
clusters. The XMM-LSS survey will serve as a valuable data base for
the identification of clusters found in the AMiBA deep and medium
surveys. We can essentially set a lower limit to the redshift of any
unidentified cluster found in the AMiBA surveys. Deeper targeted
optical and X-ray observations can then be expected to identify many
of those unidentified AMiBA clusters.

\section{Conclusions}
The next generation of SZ telescopes, such as AMiBA, can detect
clusters down to mass limits of $\sim 10^{14}$\,M$_{\odot}$ up to the
epoch of cluster formation. At last these telescopes will use the SZ
effect to produce catalogues of thousands of clusters, nearly 2 orders
of magnitude more than have previously been detected by the SZ
effect. The combination of AMiBA surveys and large optical and X-ray
surveys such as the XMM-LSS survey provides an exciting opportunity to
conduct detailed measurements of cosmological parameters and studies
of cluster physics. AMiBA is expected to be operational by 2004.

\acknowledgements

I would like to thank the AMiBA team, Mark Birkinshaw, Marguerite
Pierre and Ivan Valtchanov for valuable contribution. AMiBA is funded
by the Ministry of Education and National Science Council in Taiwan.

\end{document}